\documentstyle[11pt,aaspp4]{article}

\lefthead{Hora \& Latter}
\righthead{Near-IR Structure of Hubble 12}

\def\h2{H$_2$}
\def\s1{$v = 1 \to 0$ S(1)}

\def\J{{\it J}}
\def\H{{\it H}}
\def\about{$\approx$}

\def\Brg{Br$\gamma$}
\def\K{{\it K}}
\def\bd30{BD+30$^{\circ}$3639}
\def\etal{et al.\ }
\def\eg{e.g.,\ }

\def\cm3{cm$^{-3}$}
\def\kcm3{{\rm cm}^3~{\rm s}^{-1}}
\def\fcm3{{\rm cm}^{-3}}

\begin{document}

\title{A Butterfly in the Making:
Revealing the Near-Infrared Structure of Hubble 12}
\author{Joseph L. Hora}
\affil{Institute for Astronomy, 2680 Woodlawn Drive, Honolulu, HI 96822}
\author{William B. Latter}
\affil{NASA/Ames Research Center, MS 245-3, Moffett Field, CA  94035}

\begin{abstract}
We present deep narrowband near-IR images and moderate resolution
spectra of the young planetary nebula Hubble 12.  These data are the
first to show clearly the complex structure for this important
planetary nebula.  Images were obtained at $\lambda = 2.12$, 2.16, and
2.26 \micron.  The $\lambda = 2.12$ \micron\ image reveals the bipolar
nature of the nebula, as well as complex
structure near the central star in the equatorial region.  The images show
an elliptical region of emission which may indicate a ring or a cylindrical
source structure.  This structure is possibly related to the mechanism
which is producing the bipolar flow.  The spectra
show the nature of several distinct components.  The central object is
dominated by recombination lines of \ion{H}{1}\ and \ion{He}{1}.  The core is
not a significant source of molecular
hydrogen emission.  The east position in the equatorial
region is rich in lines of ultraviolet--excited fluorescent \h2.  A
spectrum of part of the central region
shows strong [\ion{Fe}{2}] emission which might
indicate the presence of shocks.
\end{abstract}


\section{Introduction}
The planetary nebula (PN) Hubble 12 (Hb 12; PN G111.8--02.8) has been
notable primarily because it represents one of the clearest cases
known of fluorescent molecular hydrogen emission, first determined by
Dinerstein et al.\ (1988) and recently confirmed by Ramsay et al.\
(1993).  The observed \h2\ line ratios were found to closely match
those calculated by Black \& van Dishoeck (1987) for the case of pure
fluorescent emission.  Objects which exhibit pure fluorescent spectra
have proven to be rare, since in regions with strong UV fields and
densities {\it n}\ $\gtrsim 10^4$ cm$^{-3}$, the line ratios are driven
to values which are characteristic of ``collisional'' fluorescent
emission (Sternberg \& Dalgarno 1989).  Shocks are common in PNe, and
in most PNe observed to date in which \h2 emission has been detected
the \h2 appears to be purely shock-excited or has a significant
shock-excited contribution (e.g., Beckwith \etal 1978; Storey 1984;
Latter \etal 1993; Aspin \etal 1993; Graham et al. 1993; Kastner et
al. 1994; Hora \& Latter 1994).

Radio continuum images taken at the VLA have shown Hb 12 to be a
bipolar nebula, aligned roughly north--south (Bignell 1983).  However,
these images showed only parts of the faint bipolar lobes, and no
detail of the central region was resolved.  Miranda \& Solf (1989)
obtained long-slit spectroscopy and confirmed the bipolar nature of
the source.  Their observations showed the nebula to be oriented such that
the north lobe is receding and the south lobe approaching.
In the course of their spectral
observations, Dinerstein et al. (1988) mapped the central region and
found the 2.12 \micron\ emission to be distributed in a
``doughnut-like" shell with inner and outer dimensions of $r =$
4\arcsec\ and 8\arcsec.  Such structures in the equatorial regions of
bipolar PNe might be important in understanding the process of the
onset of asymmetry and the shaping of PNe by stellar winds.  We have
therefore obtained deep \h2\ images and spectra of Hb 12 to further
probe the nature of this important and unusual nebula.

\section{Observations and Data Reduction}
Near-IR images of Hb 12 were obtained during 1995 January 21--22 UT
with the University of Hawaii (UH) QUick InfraRed Camera (QUIRC) at
the UH 2.2m telescope on Mauna Kea.  This is a new instrument which
utilizes a 1024$\times$1024 pixel ``HAWAII'' array (Kozlowski \etal
1994).  The telescope was used in the f/10 configuration, giving a pixel
size of 0\farcs18 pixel$^{-1}$.
The camera currently holds eight filters manufactured by Barr
Associates, including their narrowband (\about 1\%) filters at
$\lambda = 2.12$ and 2.16 \micron, and a 2.26 (2.7\%) \micron\ continuum
filter.  The image presented here was constructed of separate
exposures which were individually sky-subtracted and
flat-fielded before shifting and averaging.
The images are shown in Figures 1 and 2.
The average FWHM of the point sources in
the 2.12 \micron\ image is 0\farcs71, and the 1 $\sigma$ noise levels
are 7$\times 10^{-5}$, 1.1$\times 10^{-4}$, and 5$\times
10^{-5}$ Jy arcsec$^{-2}$ for the 2.16, 2.26, and 2.12 \micron\ images,
respectively.  Flux calibration was
performed using the standard star HD 3029, which was assumed to have a
magnitude of 7.09 in each of the narrow band filters, based on the K-band
magnitude of Elias \etal (1982).

Figure 2b shows a continuum-subtracted image at 2.12 \micron.  The
subtraction was performed by using the 2.26 \micron\ image, scaled
by a factor obtained by comparing the continuum levels at 2.26 and 2.12
\micron\
in the spectrum of the 3\farcs7 E position (see below).  The 2.26 \micron\
image has higher noise than the original 2.12 \micron\ image, so the
continuum-subtracted image is slightly degraded.  By scaling the
continuum image to the extended emission spectrum, we have not matched the
continuum level of the core.  Therefore we have masked the central
$\sim$3\arcsec\ where the subtraction is invalid.   A slight mismatch of
the PSF between the two wavelengths is visible in the small artifacts around
each of the field stars.  Furthermore, there is some contribution to the
flux in the 2.26 \micron\ image from some lines of \h2, primarily from the
(2,1) S(1) line at 2.248 \micron.  This is probably contributing to the
extended
emission visible in the 2.26 \micron\ image in Figure 1b.  However, the
continuum-subtracted
image shows that at 2.12 \micron\, line emission from \h2 dominates and that
it is extended in the equatorial region and in the bipolar lobes.

The spectroscopic observations were performed on 1994 September 25 UT on the
UH 2.2m telescope using the near-IR spectrometer KSPEC (Hodapp et al. 1994).
The instrument was configured with a 1.0$\times$6.5 arcsec slit, providing a
resolving power of $\lambda / \Delta\lambda \sim 700$.  Images were taken
simultaneously with the spectral integrations using the slit-viewing detector,
allowing for accurate placement and guiding.  Integrations were done at two
slit positions, one centered on the bright nebular core, and the other at a
position 3\farcs7 east, with the long dimension of the slit oriented N-S.
The extracted length of the slit was 2\arcsec\ for the core, 2\farcs5 for
the 3\farcs7 E region, and 2\arcsec\ for the 3\farcs7 E 2\arcsec S region (see
discussion of regions below).
Alternating source and sky integrations were taken
and differenced to remove sky and telescope background flux.  The on-source
integration times were 150 s on the core and 1260 s for the off-core positions.
The data were reduced and calibrated as described in Hora \& Latter (1994),
using the star SAO 35265 for atmospheric correction and HD 44612 as a flux
standard.  The 0.84 -- 2.43 \micron\ spectrum was extracted for the core, and
the 1.15 - 2.43 \micron\ spectra were extracted for the off-core positions.
The spectra of Hb 12 are shown in Figures 3 and 4, and a list of
the lines detected is given in Table 1.  The absolute calibration is uncertain
to 15-20\%, but the measurement of relative line flux and line ratios is much
more accurate.  The estimated 1$\sigma$ noise in the line flux is $\sim
3\times10^{-15}$ erg cm$^{-2}$ s$^{-1}$ for the core spectrum and $\sim
1\times10^{-16}$ erg cm$^{-2}$ s$^{-1}$ for the off-core positions.

\section{Results and Discussion}
The image shown in Figure 2 reveals that the nebula has
faint bipolar lobes and an elliptical equatorial
region with streamers, along with a compact core.  There is a faint halo
surrounding the equatorial region that might be associated with
it or the bipolar
lobes.  There is an intersection of features in the equatorial region
at the position
$\sim$4\arcsec\ east of the core where a local maximum of \h2 emission
exists.  That is the position near where the observations of
Dinerstein \etal\ (1988), Ramsay \etal\ (1993), and one of the spectra
presented here in Figure 4 were centered.

\subsection{The core}
The core region of the 2.12 \micron\ image
(Figure 2) has a FWHM size of 0\farcs72$\pm$0\farcs03,
indistinguishable from the average stellar FWHM of
0\farcs71.  It is similarly compact at the other observed wavelengths.
 This is consistent with previous optical and radio
observations which showed the nebula to be ``stellar" or having a
spatial extent of $\lesssim$ 1\arcsec~ (Johnson \etal 1978; Kohoutek \&
Martin 1981; Bignell 1983; Miranda \& Solf 1989).  The spectrum of the
core shown in Figure 3 is typical of many PNe, dominated by lines of
\ion{He}{1}~ and \ion{H}{1}.  There is good agreement between the spectrum
in Figure 3 and the spectra presented by
Rudy et al. (1993) and Kelly \& Latter (1995) where the wavelength
ranges of the measurements
overlap.  For example, we measure a Pa$\beta$ flux of
1.36$\pm0.20\times10^{-11}$ erg cm$^{-2}$ s$^{-1}$, whereas Rudy et
al.  obtained 1.13$\times10^{-11}$ erg cm$^{-2}$ s$^{-1}$ in a
9\farcs6 $\times$8\farcs1 elliptical aperture, and Kelly \& Latter
found 1.8$\times10^{-11}\ \pm\ 20$\% erg cm$^{-2}$ s$^{-1}$ in a
6\arcsec\ circular aperture.  From the \ion{O}{1}\
$\lambda = 1.1287$ to 1.3165 \micron\  line ratio (=0.3), Kelly \&
Latter (1995) argue that the \ion{O}{1}\ emission is UV continuum pumped with
no resonant pumping from Ly$\beta$. Our $\lambda = 0.8448$ to 1.3165
\micron\  line ratio (= 9.3) confirms this result.

No \h2\ is detected in the spectrum of the core region,
although the brightest \h2\ line in the 3\farcs7 E spectrum would only be
detected at the 3$\sigma$ level in this spectrum.  It is likely that there is
some \h2\ emission from the direction of the core,
since there will be at least a component
due to projection through the equatorial region.  However, its relative
flux is small compared to other lines and the continuum.  For example, the
\h2/\Brg\ line ratio is $\sim$1 in the 3\farcs7 E position, but is
$\lesssim$0.03 in the core.  Therefore, the unresolved
core itself does not appear to
be a significant source of \h2 flux in Hb 12.

Rudy et al.\ (1993) reviewed estimates of the differential attenuation
towards Hb 12.  Several estimates, including the Balmer line ratios,
indicate an $E(B-V)$ of 0.85.  However, they found the Paschen lines
showed significant line-to-line departures from Case B ratios, and a
fit to their data implied an $E(B-V)$ of 0.45.  The Brackett line
ratios that we measure have relatively low line-to-line departures
from the expected Case B ratios.  However, a determination of $E(B-V)$
from our measured values of the Brackett line fluxes gives a value of
0.28 (assuming $T=10^4$ K and $N_e=10^6$ \cm3).  It is possible that
the near-IR line emission is from different regions within the core
compared to the optical lines, or that nonstandard extinction effects
are present.  Luhman \& Rieke (1995) also measure anomalous Brackett and
Brackett/Pfund line ratios, and suggest that Br10 and \Brg\ may be
optically thick.  In that case the $E(B-V)$ values calculated based on
expected Case B ratios are not valid.

\subsection{The equatorial region and bipolar lobes}
\subsubsection{Morphology}
The walls of the bipolar lobes are visible in the \h2 image (Figure 2)
extending roughly N--S with an opening angle of $\sim70^{\circ}$, and the
axis of the nebula is at a position angle $\sim 2^{\circ}$ W of N.  The
distribution of \h2 is similar to the ionized gas as traced by the VLA
images.  However, if Hb 12 is similar to other PNe such as NGC 7027
(Graham et al.\ 1993) or M\,2--9 (Hora \& Latter 1994), the \h2
emission is most likely from a region just exterior to the ionized zone
(i.e., at a greater radius from the bipolar axis),
as determined by the radio continuum emission (Bignell 1983).
The lobes are detected out to a total nebular size of
35\arcsec\ N--S.

A deconvolution of the central region was performed using the
Richardson--Lucy (RL) method implemented in the STSDAS package of IRAF.  A
star near Hb 12 in the 2.12 \micron\ image was used as the PSF.  The
result of the deconvolution is shown in Figure 5b.
Point sources in the deconvolved image
were reduced to a FWHM of $\sim$0\farcs4.  The structure visible in
the raw images is significantly enhanced by the deconvolution;
sharp edge of the equatorial region appears incomplete N and E of the core.
Brighter spots inside the edge are at the intersection of filamentary
structures, rather than in discrete clumps.
The deconvolution process does not seem to have created any artificial
structure in the extended emission since features visible in the RL image
can also be seen in the original.
The apparent exception to this is within a radius of 2 arcsec from the core.
This area may have been affected
by detector nonlinearities since the core was near saturation in the
individual exposures.   Artifacts can arise in RL deconvolved images
(e.g., Linde \& Johannsesson 1991).  We have not performed any independent
tests
of the validity and possible problems with the RL deconvolution technique.
However, none of the conclusions in this
paper depend on the accuracy of the RL image in Figure 5b.

The filaments and outer edge of the elliptical equatorial region
are the locations of UV-excited fluorescent
\h2\ emission, as seen in Figure 2 and discussed below.  Part of the
slit overlapped the edge as these data for the east position were
taken.  We have extracted these data separately and plotted them in
Figure 4 (marked as 3.7 E 2 S).  This region appears to have roughly
similar \h2\ line ratios compared to the 3\farcs7 E
position. However, strong lines of [\ion{Fe}{2}] are seen at $\lambda =
1.644$ and 1.257 \micron\ that are not present or relatively weaker in the
core and 3\farcs7 E positions.  These strong features indicate that a
shocked region might be present in the edge.  The edge  structure might
represent a
transition between the relatively low density region inside and higher
density outside.  The outer edge of the bipolar lobes
extend N-S from the equatorial region, indicating some relationship between the
equatorial region and the formation of the lobes.
The width of the edge emission is not resolved in these images.

There are several possible structures which could explain the observed
morphology.  One is a ring or torus that encircles the core in a plane
perpendicular to the roughly N--S bipolar axis.   This was suggested by
Dinerstein et al. (1988) who found the \h2 emission to be distributed in
an elliptical region around the core.  The outer edge of the equatorial
zone as shown in Figure 2 is roughly elliptical, with a projected size of
12\farcs5$\times$6\farcs9, and assuming the ring is circular, this
implies the ring plane is tipped $\sim 30^{\circ}$ from the line of sight.

Another possibility is that the observed ``ring" structure is
significantly distorted from circular, or that the structure is not
an equatorial ring but a shell.   This is suggested
by the fact that the observed structure does not precisely match what
one expects from a true ring.  In particular, the extreme E and W
positions come to a sharp point or cusp rather than being more rounded
as one would expect from a torus.  This
is illustrated in Figure 6a, where a tilted ring has been superimposed
on an image of the inner region.  Also, the brightness of the emission
at the E and W ends is less than or equal to the other parts of the
ring.  One might expect that these regions would be slightly
limb-brightened.

An alternate model of the structure of Hb 12 is a short cylindrical
shell.  This has been proposed for several other young PNe, such as NGC
7027 (Scott 1973) and \bd30 (Bentley et al. 1984).  Recent two-dimensional
radiation--gasdynamic simulations by Frank \& Mellema (e.g., 1994a)
have shown that such structures can be generated by interacting winds
with the proper initial conditions.  One characteristic of these
models is that they reproduce the ``eye'' appearance as the result of
projection of the top and bottom circular edges of the cylinder seen
at moderate inclination angles.  Figure 6b shows the outlines of a
cylinder on the central region of the nebula.  The inclination angle
of the cylinder axis is 50$^{\circ}$ from the plane of the sky.  There
is \h2\ emission from other parts of the central region that roughly
outline the cylindrical shape, but the brightest emission is from the
overlapping regions of the ends of the cylinder,
looking through what would be the open end.
This could be caused by absorption of
the emission from the other parts of the cylinder by the material
outside of the inner region.  The presence of absorbing material is
also suggested by the fact that the bipolar lobes that extend to the N
and S are brightest at some point around 10\arcsec\ from the core, not
at their closest point.  High-resolution spectra of the \h2\ emission
to determine the velocity structure of the equatorial region and filaments
will be necessary to better understand the observed morphology.

\subsubsection{Spectra}

The spectrum of the 3\farcs7 E position in Figure 4 is similar to
previously published spectra (Dinerstein et al. 1987, Ramsay et
al. 1993) but has several significant differences.  For example, the
ratio of \Brg\ to \h2 \s1 line emission is $\sim$2 in Ramsay et al.'s
spectrum, whereas we find the ratio to be close to 1.  A similar
discrepancy is seen for the lines of \ion{He}{1}\ at $\lambda = 2.058$ and
2.113 \micron.  This can perhaps be explained by the different slit
sizes used.  Ramsay et al.\ summed spectra over a
3\farcs1$\times$15\farcs5 region, whereas the spectrum for the
3\farcs7 E position presented here is summed over a region
1\arcsec$\times$2\farcs5.  The Ramsay et al. spectrum therefore
includes part of the equatorial region edge, and probably a contribution from
the core or regions near the core where the \h2 \s1 line is not
detected.  There also might be significant scattered emission from the
central source at the eastern positions.  We also detect about 30
lines in the \J\ and \H\ bands, and the lines seen in the \K\ band by
Ramsay et al.\ plus the following additional lines have been detected:
$v = 1 \to 0$ S(3) (1.9570 \micron), 8 -- 6 O(2) (1.9702 \micron), 7
-- 5 O(5) (2.0215 \micron), 3 -- 2 S(5) (2.0650 \micron), 9 -- 7 Q(2)
(2.0835 \micron), 9 -- 7 Q(3) (2.1001 \micron), 3 -- 2 S(4) (2.1274
\micron), and 9 -- 7 O(3) (2.2530 \micron).

\subsection{\h2 excitation (Revisited)}

Molecular hydrogen emits a near-IR spectrum through emission from slow
electric quadrupole vibration--rotation transitions. There are two
mechanisms by which these transitions can be excited. Collisional
excitation can occur in gas with kinetic temperatures $T_K\ \gtrsim
1000$ K (most often associated with moderate velocity shock
waves). \h2\ can also be excited by absorption of ultraviolet photons
with $\lambda > 912$ \AA\ in the Lyman and Werner bands.  In low
density regions ($n_{\rm tot} < 10^4 \fcm3$) the resulting cascade
produces an easily identifiable near-infrared spectrum (Black \& van
Dishoeck 1987).

Because of its clear spectral signature of UV-pumped \h2, Hb 12 has
become the prototype source of near-IR fluorescent \h2\
emission. Since regions displaying this type of extreme spectral
signature appear to be relatively rare, it is worthwhile to examine
the spectrum of molecular hydrogen from Hb 12 is some detail. Our
spectra sensitively cover most of the near-IR {\it J}, {\it H}, and
{\it K} bands and give some spatial information. We have therefore
re-examined the \h2\ excitation at two points in the nebula
associated with a ``knot'' of \h2\ 3\farcs7 East of the
core, and a point on the edge of the equatorial
region 3\farcs7 East and 2\arcsec\
South of the core.  Luhman \& Rieke (1995) discuss a longslit spectrum
taken in the East--West direction. Their study provides data that
overlaps with ours, provides information on regions in the nebula not
previously observed, and goes into greater detail than will be done
here.

In the spectrum of the position 3\farcs7 East of the core, $\approx
50$ lines of \h2\ and $\approx 35$ lines in the 3\farcs7 E, 2\arcsec S
position are detected (Fig.\ 4 and Table 1).  These lines originate
from vibrational levels that range from $v = 1$ to $v = 9$. We are,
therefore, able to extend the studies of Dinerstein et al.\ (1988) and
Ramsay et al.\ (1993). The following analysis follows that of Ramsay
et al. (1993) for Hb 12, and that of Hora \& Latter (1994) as applied
to M 2--9 and AFGL 2688.

The long spectral baseline of these data allows for a good
determination of the differential attenuation in the wavelength region
observed. For the spectrum at 3\farcs7 East, we used 8 line ratios
from 16 different transitions that extend from $\lambda = 1.167$ to
2.423 \micron. Six line ratios of the type needed to determine
attenuation are available in the 3\farcs7 E, 2\arcsec\ S spectrum.
Each ratio was for a line pair originating from the same upper state,
making the predicted ratio just the ratio of the products $\Delta E
A_{ul}$ for each of the transitions. Several ratios are clearly
discrepant, either because of measurement error, or contamination by
other emission lines.  These ratios were not used in the attenuation
determination. Using a standard interstellar extinction law (Rieke \&
Lebofsky 1985), the visual attenuation toward both positions is
estimated to be $A_V \approx 3.7^{+1.5}_{-1.0}$ mag.  In all computations
presented here, a value of $A_V
= 3.5$ mag has been used to deredden the observed \h2\ line
ratios. In the type of analysis that follows, any errors introduced by
departures from a purely interstellar type extinction law, or simply
from uncertainty in the value of $A_{V}$ will be small.

A test of whether the kinetic temperature is high enough to excite the
IR emission is the \h2\ ortho-to-para ratio. A lower than thermal
ratio (the ratio of statistical weights = 3/1) indicates temperatures
and densities much lower than are required to collisionally excite the
near-IR spectrum. O/P interchange is accomplished through reactions
with H$^+$ and H$^+_3$ (for example). This is in competition with the
\h2\ source reaction (grain surface reactions), which is thought to
produce \h2\ initially in the thermalized 3/1 ratio. Ramsay \etal\
determined an uncertain ortho-to-para ratio of 1.72 based only on a
few lines in the $K$ band. By a similar analysis that used many more
line pairs, we find the ortho-to-para ratio to be $1.73\pm 0.2$, in
excellent agreement with the earlier result.  This O/P ratio therefore
indicates that the \h2\ emission is not thermally excited.

In Figure 7 the logarithm of the observed populations (using an
ortho-to-para ratio = 1.73) is shown relative to that in the $v=1,\
J=3$ level (the origin of the \s1 line) versus $T_{upper} =
E(v^\prime, J^\prime)/k$. All points would lie on a single straight
line if a strict Boltzmann distribution held for the
vibration--rotation levels (see, \eg the case for AFGL 2688 in Hora \&
Latter 1994).  It is clearly evident in Fig.\ 7 that no such straight
line exists. However, each set of points from common vibrational
levels fall on individual lines for which a single slope applies. The
negative inverse of this slope gives the rotational excitation
temperature of the molecules along the line of sight. The average
rotational excitation temperature found from linear regression to all
the points is $T_{ex}(J) = 1395\pm 450$ K. A vibrational excitation
temperature can be determined by comparing lines arising from the same
rotational level. We have done this for 27 line pairs and found that
no single temperature describes the system. In a general way, the
average excitation temperature tends to increase with vibrational
level and typically falls in the range $T_{ex}(v) \sim 7000 - 13500$
K. These types of excitation temperatures are characteristic of a pure
fluorescence spectrum.  While the signal-to-noise ratio in the
3\farcs7 E 2\arcsec\ S spectrum is lower, thereby increasing the
uncertainties, the same analysis of that spectrum proves that the
3\farcs7 E 2\arcsec\ S  region
has a purely UV excited spectrum as well. We find no clear difference
in excitation characteristics for the two locations. If shocks
contribute to the excitation of the \h2\ within the equatorial region,
there is no indication of it in the data presented here.

A comparison of the spectrum at the 3\farcs7 E position with the
models of Black \& van Dishoeck (1987) was made by directly comparing
the observed line ratios with those predicted by their Model 14
($n_{\rm H} = 3000\ \fcm3$ and UV field 1000 times the ambient
interstellar field). After removal of lines that are known to be
contaminated, or are likely to be contaminated by nearby atomic lines,
the agreement is excellent. All lines originating from $v=1$ and $v=2$
levels typically agree to within 10 to 20\%. Discrepancies increase
for $v \geq 3$ to a maximum factor of $\sim 2.5$ for 3 lines, but all
others fall within a factor of 2.0 of the predicted ratios. An even
better fit likely could be found by optimizing the density and UV
field parameters.

\subsection{A possible link between bipolar reflection nebulae and
butterfly PNe}

In many respects, the new \h2\ image of Hb 12 in Figure 2 is strikingly similar
to that of the more evolved butterfly nebula NGC 2346 (\eg Latter
\etal 1995). By comparing these two objects and looking at how Hb 12
compares with younger bipolar nebulae, perhaps some new light can be
cast on the evolution of axisymmetric PNe. It is in this area that the
distinction of whether or not a thin torus is present can be of some
significance.

Hb 12 is clearly in a period of very rapid evolution
(Luhman \& Rieke 1996). Other PNe in a
similar state might be NGC 7027 and \bd30. Neither of these objects
display fluorescent \h2\ emission to the degree found for Hb 12. Given
that both have sufficiently strong UV emitting central stars, this
likely argues for higher density envelopes, and therefore (perhaps)
more massive progenitors with high mass loss rates on the AGB. Those
objects, while not spherically symmetric, also do not appear as
extremely asymmetric as Hb 12. For these reasons it is difficult, and
perhaps not justified to make any direct comparisions with other
objects in short periods of rapid evolution.

Even though it is a very young extreme butterfly nebula, Hb
12 has already established features that are directly comparable to
more evolved objects like NGC 2346. Also, as a relatively young
PN, it might allow us to say something about the morphology of the
post-AGB bipolar nebulae. As with Hb 12, the ``wings'' of NGC 2346 are
seen predominantly in \h2\ emission. There is also a central
elliptical region of filamentary \h2\ emission, not unlike that seen
in Hb 12. The \h2\ emission observed in NGC 2346 is likely from shocked
gas at the interface of expanding bubbles (Kastner \etal 1994, Latter \etal
1995), which is unlike Hb 12 where the \h2\ emission arises from
radiative effects in the photodissociation region (PDR). From the data
available so far, we cannot tell what the excitation mechanism is for
the \h2\ ``wings'' in Hb 12.

Hb 12 and NGC 2346 likely have evolved from post-AGB objects like AFGL
618 and AFGL 2688. These objects are bipolar reflection nebulae, in
which the density of dust decreases with increasing stellar latitude
(Morris 1981; Yusef-Zadeh \etal 1984; Latter \etal 1992; Latter \etal
1993). Molecular hydrogen emission is detected in the lobes of those
objects and others like them.  AFGL 618 and M 2--9 are known to be at
a phase in which at least part of the \h2\ emission is coming from
fluorescent excitation within the PDR (Latter \etal 1992; Hora \&
Latter 1994). It is possible that we are seeing an evolutionary sequence
in these objects
-- from young to old. It is difficult to know for certain, given the
wide range of possible central star and envelope masses. However, the
similarities imply a strong relationship. All have a bipolar symmetry
and all have \h2\ emission arising, at least in part, from the bipolar
lobes. These observations are consistent with the lobes being
expanding bubbles with the \h2\ emission coming from limb-brightened
edges.

It has been argued that compact, overdense tori (with a density in
excess of what would be deposited by heavy mass loss
on the AGB) are not required to
explain the observed morphologies of the young bipolar nebulae
(\eg Morris 1981; Yusef-Zadeh \etal 1984; Latter \etal
1993). There are exceptions -- objects such as AFGL 915 and possibly
IRAS 09371+1212 appear to have dense, long-lived disks (Jura \etal
1995; Roddier \etal 1995). However, for the other objects there is a
preferred direction for the mass loss resulting in a lower dust
density in the polar regions -- it is this lower dust density that
produces the observed bipolar shapes in scattered starlight. The
central regions of Hb 12 and NGC 2346 are complex, but
there is not clear evidence for the remnant of a dense torus of the
type required to be the primary mechanism of bipolar shaping. What we
find are structures apparently formed by instabilities in interacting winds in
axisymmetric distribution of material. This is what would be expected
from the above mentioned general morphology of bipolar nebulae, and
from a number of recent radiation--gasdynamics models (\eg Mellema
1993; Frank \etal 1993; Frank \& Mellema 1994a,b).  While large
uncertainties remain, Hb 12 might be an important link between the
young bipolar nebulae and the evolved butterfly PNe.

\section{Conclusions}

Near-IR images and spectra of Hb 12 reveal the complex structure of
the inner region of this PN, and its large-scale bipolar
morphology. UV flux from the bright, unresolved core excites
fluorescent emission from \h2\ in the equatorial region.  The \h2\
fluorescence is consistent with low-density ($n < 10^4$ \cm3) gas in
this region.  The presence of apparently shocked [\ion{Fe}{2}] emission in
the edge of the equatorial region could signal a transition to higher
density material outside
of this zone.  The images suggest a toroidal or cylindrical structure which
delineate an equatorial density enhancement which is channeling the bipolar
flow.  The structure compares well with recent radiation-gasdynamic model
results that show similar morphologies in early PNe development.

We have re-examined the \h2\ emission from Hb 12 using new spectral
and spatial information. For the 3\farcs7 E position, linear fits to
the different vibrational levels finds the rotational excitation
temperature to be $T_{ex}(J) = 1395\pm 450$ K, and vibrational
excitation temperatures typically fall in the range $T_{ex}(v) \approx
7000 - 13500$ K, but no single vibrational temperature describes the
excitation.  Excellent agreement is seen between the observed line
ratios and those predicted by theoretical \h2\ fluorescence
calculations.

Advances in near-infrared astronomy have helped to provide a good
sampling of objects in transition across the top of the HR
diagram (Kwok 1996). The goal of providing a clear evolutionary sequence is
not yet possible.  Any evolutionary sequence of
observed objects will be highly uncertain because of extreme
differences in evolutionary timescales for objects of differing
mass. Continued observations and theoretical modeling are
required to establish this connection.
However, with Hb 12 now revealed as a possible link between the
young bipolar reflection nebulae and the more evolved butterfly PNe, a
full description of the evolution of axisymmetric nebulae might be closer at
hand.

\acknowledgments

We wish to express thanks to Lynne Deutsch for her assistance in
obtaining the spectral data, and discussion on drafts of the paper. We
thank John Black for illuminating conversations, and Kevin Luhman for
communicating results prior to publication.

\newpage
%
%
\begin{figure}
\caption{The planetary nebula Hubble 12.  North is up and east
to the left in all images.
a)  Image taken with the \Brg\ filter at 2.16 \micron.  b)  Continuum
(2.26 \micron\ filter). }
\end{figure}
%
%
\begin{figure}
\caption{The planetary nebula Hubble 12 at 2.12 \micron.  North is up and east
to the left in all images.
a)  2.12 \micron\ image of Hb 12.
b)  Continuum-subtracted image, obtained by subtracting a scaled 2.26 \micron\
image from the 2.12 \micron\ image in 2a).  The central $\sim$3\arcsec\ has
been masked where the continuum level is not well matched.}
\end{figure}
%
%
%
\begin{figure}
\figcaption{The spectra of Hubble 12, at the core position.
The wavelength positions of lines of
\protect \ion{H}{1}, \protect \ion{He}{1},
and other strong lines are indicated.  Note that the
wavelength scales for each of the ranges are slightly different.  The
apparent emission features between 2.0 and 2.02 \micron\ are due to
differences in the atmospheric absorption when the reference star data were
taken.
}
\end{figure}
%
%
%
\begin{figure}
\figcaption{The spectra of Hubble 12 at the 3\farcs7 E
0\arcsec\ N and 3\farcs7 E 2\arcsec\ S positions, shown with the core spectrum
for comparison.
The wavelength positions of lines of
\protect \ion{H}{1}, \protect \ion{He}{1},
H$_2$, and other strong lines are indicated.
}
\end{figure}
%
%
%
\begin{figure}
\caption{
The inner 20\arcsec\ of Hubble 12 at 2.12 \micron.
a) The original data image (Figure 2) expanded
and rescaled to show the structure of the equatorial region.  b) Richardson-
Lucy deconvolution of the image in a).  The FWHM of point sources has been
reduced to $\sim$0\farcs4 in this image.
}
\end{figure}
%
%
%
\begin{figure}
\caption{
The inner 20\arcsec\ of Hb 12 at 2.12 \micron.  a) Overlayed with a circle
viewed at
$\sim$30\arcdeg\ from its plane.
b) Overlayed with an outline of a short cylinder,
its axis tilted $\sim$50\arcdeg\ from the plane of the sky.
}
\end{figure}
%
%
%
\begin{figure}
\caption{Excitation diagram for Hb 12 in the 3\farcs7 E position (dereddened
for $A_V = 3.5$ mag). Shown are the upper state vibration-rotation populations
relative to that in the $v=1,\ J=3$ level versus the energy of the upper state
in Kelvin. Here $g_J$ is the statistical weight, which for odd $J$ levels
includes the ortho-to-para ratio of 1.73 (see text). Linear fits to the
different vibrational levels finds the rotational excitation temperature to be
$T_{ex}(J) = 1395\pm 450$ K. The lines shown are characteristic for this value
of $T_{ex}(J)$. Ratios not used in the analysis because of blending are not
plotted.}
\end{figure}

\end{document}